\documentclass[aps,pra,twocolumn,citeautoscript]{revtex4-1}

\usepackage{amsmath,amssymb}
\usepackage{bm}
\usepackage{graphicx}
\usepackage{epstopdf}
\usepackage{latexsym}
\usepackage{subfigure}
\usepackage{color}
\usepackage{dsfont} 
\usepackage{wasysym} 
\usepackage{stmaryrd} 
\usepackage{hyperref} 

\definecolor{airforceblue}{rgb}{0.36, 0.54, 0.66}

\newcommand{\be}{\begin{equation}}
\newcommand{\ee}{\end{equation}}
\newcommand{\bea}{\begin{eqnarray}}
\newcommand{\eea}{\end{eqnarray}}

\usepackage{tikz}

\def\plaquettev{\tikz[baseline=.05ex]{
\fill (0,0) circle (1pt) coordinate (A);
\fill (1.5ex,0) circle (1pt) coordinate (B);
\fill (2.25ex,1.3ex) circle (1pt) coordinate (C);
\fill (1.5ex,2.6ex) circle (1pt) coordinate (D);
\fill (0ex,2.6ex) circle (1pt) coordinate (E);
\fill (-0.75ex,1.3ex) circle (1pt) coordinate (F);
\fill (-0.375ex,-0.65ex) circle (0pt) coordinate (AA);
\fill (1.875ex,-0.65ex) circle (0pt) coordinate (BB);
\fill (3.0ex,1.3ex) circle (0pt) coordinate (CC);
\fill (1.875ex,3.25ex) circle (0pt) coordinate (DD);
\fill (-0.375ex,3.25ex) circle (0pt) coordinate (EE);
\fill (-1.5ex,1.3ex) circle (0pt) coordinate (FF);
\draw (A)--(B);
\draw [ultra thick] (B)--(C);
\draw (C)--(D);
\draw [ultra thick] (D)--(E);
\draw (E)--(F);
\draw [ultra thick] (F)--(A);
\draw [ultra thick] (A)--(AA);
\draw [ultra thick] (B)--(BB);
\draw [ultra thick] (C)--(CC);
\draw [ultra thick] (D)--(DD);
\draw [ultra thick] (E)--(EE);
\draw [ultra thick] (F)--(FF);}
}

\def\plaquetteh{\tikz[baseline=.05ex]{
\fill (0,0) circle (1pt) coordinate (A);
\fill (1.5ex,0) circle (1pt) coordinate (B);
\fill (2.25ex,1.3ex) circle (1pt) coordinate (C);
\fill (1.5ex,2.6ex) circle (1pt) coordinate (D);
\fill (0ex,2.6ex) circle (1pt) coordinate (E);
\fill (-0.75ex,1.3ex) circle (1pt) coordinate (F);
\fill (-0.375ex,-0.65ex) circle (0pt) coordinate (AA);
\fill (1.875ex,-0.65ex) circle (0pt) coordinate (BB);
\fill (3.0ex,1.3ex) circle (0pt) coordinate (CC);
\fill (1.875ex,3.25ex) circle (0pt) coordinate (DD);
\fill (-0.375ex,3.25ex) circle (0pt) coordinate (EE);
\fill (-1.5ex,1.3ex) circle (0pt) coordinate (FF);
\draw [ultra thick]  (A)--(B);
\draw (B)--(C);
\draw [ultra thick]  (C)--(D);
\draw (D)--(E);
\draw [ultra thick]  (E)--(F);
\draw (F)--(A);
\draw [ultra thick] (A)--(AA);
\draw [ultra thick] (B)--(BB);
\draw [ultra thick] (C)--(CC);
\draw [ultra thick] (D)--(DD);
\draw [ultra thick] (E)--(EE);
\draw [ultra thick] (F)--(FF);}
}

\begin{document}

\title{Quantum loop states in spin-orbital models on the honeycomb lattice}

\author{Lucile Savary}
\affiliation{Department of Physics, Massachusetts Institute of
  Technology, 77 Massachusetts Ave., Cambridge, MA 02139}

\date{\today}
\begin{abstract}
We construct a physically realistic and analytically tractable model for spin-1 systems with
orbital degeneracy on the honeycomb lattice, relevant to honeycomb materials
with large Hund's and weak spin-orbit couplings, and two electrons
in $t_{2g}$ orbitals. This model realizes many new phases whose building
blocks are orbital loops decorated by Haldane chains. These include a Haldane loop crystal, a symmetry-protected topological
phase, and, notably, a regime where the decorated loops resonate. When
taken to the three-dimensional hyperhoneycomb lattice, the latter
regime becomes a (symmetry-enriched) $U(1)$ quantum spin-orbital liquid, ``disordered'' both in the spin and orbital
channels. We hope this construction will pave the way for realizing
many of the Haldane-chain-based phases which have been theoretically
proposed in the literature. 
\end{abstract}

\maketitle

The wide variety of proposals for exotic ground states of many-body
Hamiltonians calls for physically-realistic models prone to yield such
states. Among those are, for example, quantum spin liquids (QSLs) and
interacting topological insulators (fermionic or bosonic), also known as
``symmetry-protected topological'' (SPT) phases. A very general
characteristic of these phases is the existence of fractional
excitations, either in the bulk or at the edge of the system. The original proposal of Anderson for
QSLs involved ``resonating valence bonds,'' i.e.\ coherent
superpositions of singlet coverings of the lattice \cite{anderson1973,balents2010}. More recently,
proposals for both QSLs \cite{hao2010,li2014,huang2014} and SPTs \cite{vishwanath2013,xu2013,chen2014,wang2015} have emerged which are now based on
fluctuating {\em chains} rather than singlets. More precisely, the
building blocks are Haldane-like chains \cite{haldane1983a,haldane1983b,affleck1987}, which are featureless in
their ``bulk'' but host protected gapless states confined to their
ends. This is clearest in the AKLT chain (a representative state of the Haldane phase)
\cite{affleck1987}, where each spin one is rewritten as two spin
half's subsequently projected back onto the $S=1$ representation at
each site, and singlets form astride each bond. In this picture, two
``free'' $S=1/2$ are indeed left at each end of open chains.
The Haldane states are themselves one-dimensional SPTs; in the two and
three-dimensional Haldane-based QSL and SPT constructions, their
physical supports fluctuate and their ends
act as the bulk or edge fractional excitations. While such wavefunctions and even parent Hamiltonians
have been proposed, it has remained far from obvious how they could be achieved
in a realistic setting, let alone an actual material. 

Independently,
concrete spin-orbital (``Kugel-Khomskii'' \cite{kugel1973,khaliullin2005}) models, which capture single-site spin and orbital degeneracies, have been shown to host a rich
spectrum of
phenomena \cite{khaliullin2005}, notably, valence bond solids \cite{pen1997,dimatteo2004,vernay2004,dimatteo2005,jackeli2007,jackeli2008,normand2008,kimber2014,koch2015} and orbital liquids \cite{feiner1997,normand2008,chen2009a,chen2009b,natori2015chiral,ish2015}. The 
crucial ingredient is the modulation of the
effective spin exchange strength, which allows for stronger and weaker
bonds to form, owing to the relationship between effective exchange
strength and orbital overlap.

Here, we
show that orbital degrees of freedom provide a simple loop-forming
mechanism, and allow to naturally realize the AKLT chain ground state
picture. Specifically, we construct a
spin-orbital model, i.e.\ a
model with orbital degeneracy, on the honeycomb lattice for
$S=1$ and effective $L=1$, which supports fluctuating Haldane chains
(subtended by ``orbital loops,'' i.e.\ closed strings of bonds with large
orbital overlap), a Haldane chain based SPT, as well as a hexagon Haldane
loop crystal with ``Haldane-gap wave'' excitations. When taken to the three-dimensional hyperhoneycomb
lattice, the model is also home to a fully-fledged symmetry-enriched $U(1)$ Coulombic
spin-orbital liquid and a fractionalized antiferromagnet.

\begin{figure}[htb]
  \centering
  \includegraphics[width=3.3in]{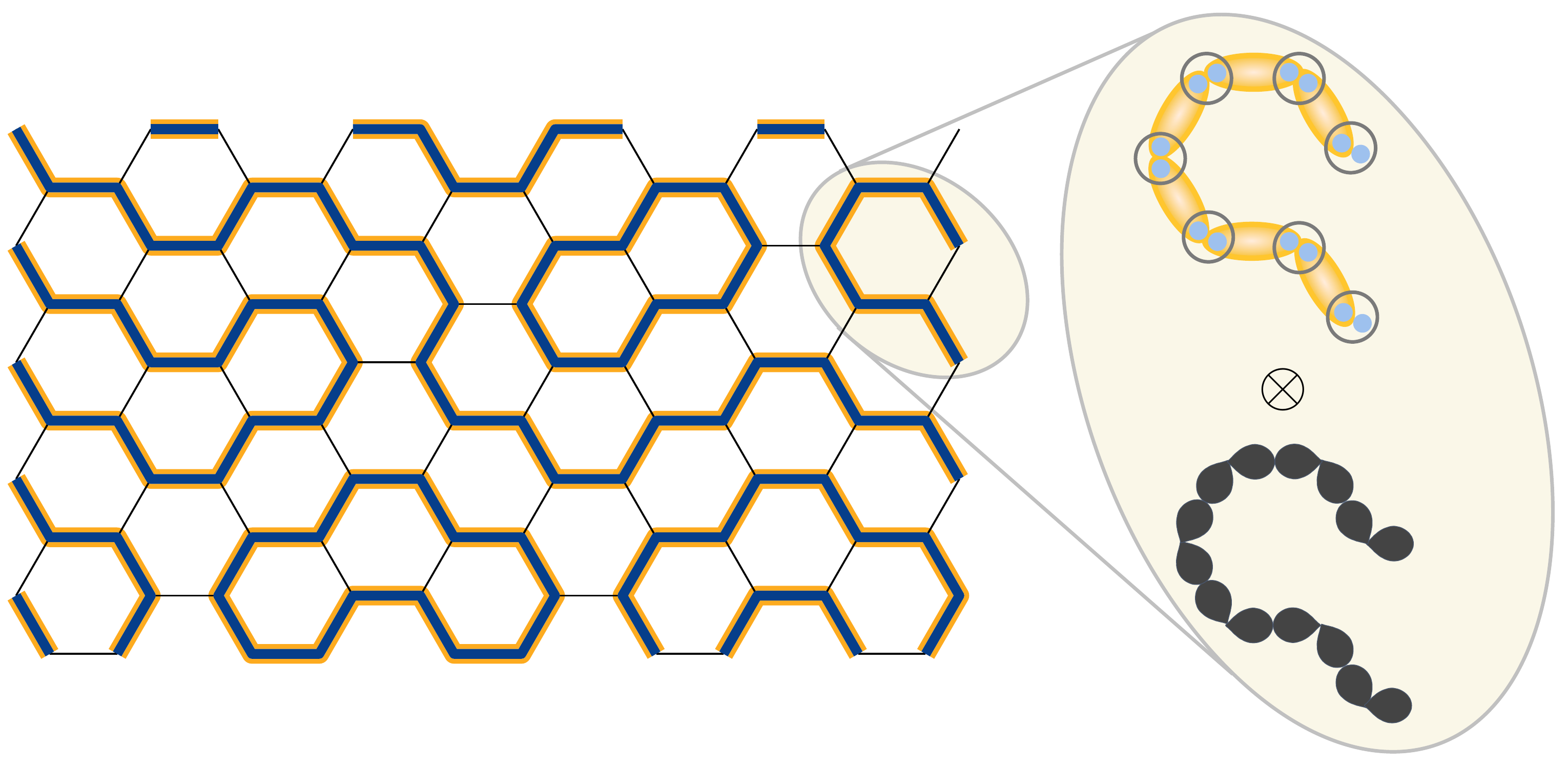}
  \caption{An orbital and Haldane loop covering of a section of the honeycomb lattice.}
  \label{fig:covering}
\end{figure}

We proceed as follows. We
first introduce the appropriate ingredients and mathematical formalism and derive the
minimal realistic model which induces the formation of fluctuating loops.
Then, we analyze in detail the pure orbital
part of the Hamiltonian and show how orbital loops emerge, before introducing spin degrees of freedom. The addition of large spin exchange produces new,
fluctuating, decorated loops. Along the way we derive results in a
large portion of the phase diagram we set out to study.

\begin{figure}[htb]
  \centering
  \includegraphics[width=3.3in]{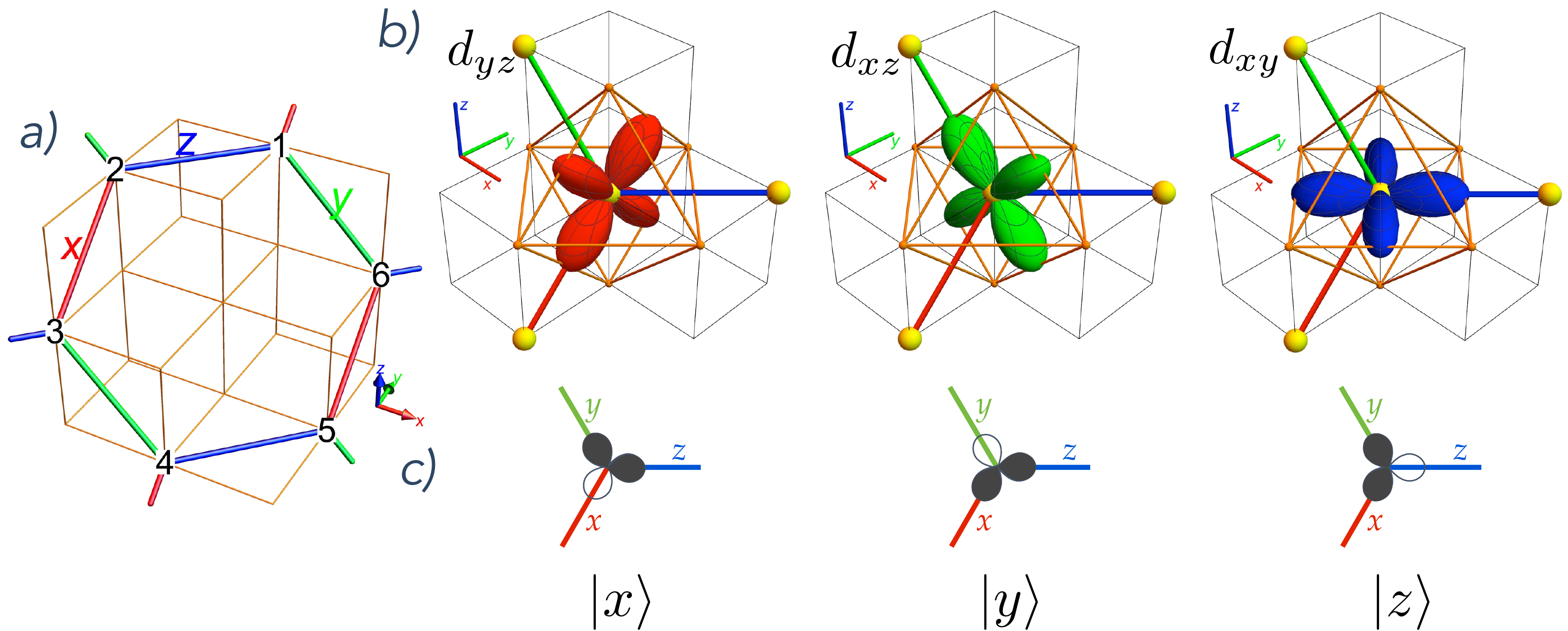}
  \caption{a) The honeycomb lattice embedded in a cubic
    structure. Honeycomb planes are perpendicular to
    $\langle111\rangle$ axes, here the $[1\overline{1}1]$ axis. b) The
    $t_{2g}$ orbitals shown in a cubic environment, surrounded by a
    putative octahedral cage. c) Pictorial representation of the
    $|x\rangle,|y\rangle,|z\rangle$ states. In
    state $|x\rangle$, the $d_{yz}$ orbital is empty, while orbitals
    $d_{xy}$ and $d_{xz}$ each contain one electron. For clarity, only the lobes in
  the bond directions are shown.}
  \label{fig:cubicenv}
\end{figure}

We consider two electrons at
each site of a honeycomb lattice, in degenerate
$t_{2g}=\{d_{yz},d_{xz},d_{xy}\}$ orbitals (which we also denote
for convenience $x$, $y$
and $z$ orbitals, respectively) \footnote{with all other orbitals filled or empty and far away in energy from
the $t_{2g}$ manifold}. We assume 
large Hund's coupling $J_{\rm H}$, which enforces the high-spin state $S=1$,
and large intra-orbital repulsion $U$, which imposes no more than one
electron per orbital. There are then two occupied and one
empty orbital at each site, and the site Hilbert space is
$\mathcal{H}=\mathcal{H}_{L_{\rm eff}=1}\times\mathcal{H}_{S=1}$. The orbital
space basis $(|x\rangle,|y\rangle,|z\rangle)$ is defined such that in
state $|x\rangle$ the
$x$ orbital is empty while the other two are filled, and
similarly for $|y\rangle$ and $|z\rangle$ (see
Fig.~\ref{fig:cubicenv}b,c)) \footnote{More formally, we can write
  that state \unexpanded{$|\mu\rangle$} is such that $d_{\mu-1,\mu+1}$
  is empty, where $x\pm1=y,z$, etc..} \cite{tchernyshyov2004}.
A set of nine 
operators acting in this space can be chosen
to be 
$\{L^\mu,P^\mu,T^\mu\}$ with $\mu=x,y,z$, such that $L^x=i(|z\rangle\langle y|-|y\rangle\langle z|)$,
$P^x=1-|x\rangle\langle x|=\hat{n}^x$ and $T^x=-(|z\rangle\langle
y|+|y\rangle\langle z|)$ and cyclic permutations. The $L^\mu$ operators are Hermitian, obey
the angular momentum algebra, and are such that
$L^\mu|\mu\rangle=0$. $P^\mu$ is a projection operator which measures
the occupation of the $\mu$ \footnote{i.e.\ $d_{\mu-1,\mu+1}$ with ($x\pm1=y,z$ and permutations)}
orbital, so that the two-electron-per-site constraint is written
$\sum_\mu P^\mu=L(L+1)=2$. Moreover, $[P^\mu,P^\nu]=0$. 

We now 
write the minimal physically realistic Hamiltonian acting in
$\mathcal{H}$,
including only nearest-neighbor interactions, which realizes a
resonating chain regime. We assume isotropy in spin space (no
spin-orbit coupling) and a local cubic environment (necessary for $t_{2g}$ orbitals). This Hamiltonian is:
\begin{eqnarray}
      H&=&\sum_{\langle
    ij\rangle}\Big(P_{i}^{\gamma_{ij}}P_{j}^{\gamma_{ij}}\left[-\zeta+J\left(\mathbf{S}_i\cdot\mathbf{S}_j+\beta(\mathbf{S}_i\cdot\mathbf{S}_j)^2\right)\right]\nonumber\\
&&\left. \qquad\qquad-\upsilon\left[T_i^{\gamma_{ij}-1}T_j^{\gamma_{ij}+1}+{\rm
   h.c.}\right]\right).
  \label{eq:1}
\end{eqnarray}
Except where otherwise noted, we take $\zeta,J\geq0$ and
$-1\leq\beta\leq1$. $\upsilon$ can always be chosen
positive, up to a gauge transformation (see Supp.\
Mat.). $\gamma_{ij}=x,y,z$ denotes the bond type of bond
$\langle ij\rangle$ (in a cubic environment each of the three
types of honeycomb bonds is orthogonal to a different
cubic axis $x,y,z$ and may be thereby labeled, see
Fig.~\ref{fig:cubicenv}), and $x\pm1=y,z$ etc.. For
example, if $\langle ij\rangle$ is a $z$-type bond, 
\begin{eqnarray}
  \label{eq:2}
 H_{\langle ij\rangle\in
  z}&=&P_{i}^{z}P_{j}^{z}\left[-\zeta+J\left(\mathbf{S}_i\cdot\mathbf{S}_j+\beta(\mathbf{S}_i\cdot\mathbf{S}_j)^2\right)\right]
  \\
&&\quad-\upsilon\left[T_i^{y}T_j^{x}+{\rm h.c.}\right].\nonumber
\end{eqnarray}
 The physical
relevance of the $\zeta,J,\beta,\upsilon$ parameters is rooted in the relative geometry of
the $t_{2g}$ orbitals and honeycomb bonds. Indeed the, e.g., $d_{xy}$
(or ``$z$''-) orbitals at each end of a $z$ bond have a large
overlap while all other overlaps are weak (see
Fig.~\ref{fig:cubicenv}); the first term in
Eq.~\eqref{eq:1} enforces precisely this concomitance of bond and orbital
filling types across a bond. All terms in Eq.~\eqref{eq:1} arise from
standard orbital-dependent superexchange mechanisms
\cite{jackeli2008,khaliullin2013}, and Eq.~\eqref{eq:1} {\em with
$\upsilon=0$ and $\zeta<0$} was studied in detail in
Ref.~\onlinecite{jackeli2008} \footnote{It was noted loop states were
  not favored for the parameter values considered.}.

We now proceed to the analysis of this model.

{\em \underline{\textbf{Orbital sector: fluctuating orbital loops.}}---}First, we set $J=0$, and investigate the
orbital part of the Hamiltonian, i.e.
\begin{equation}
  \label{eq:4}
H_{\rm orb}=\sum_{\langle
    ij\rangle}\left(-\zeta P_{i}^{\gamma_{ij}}P_{j}^{\gamma_{ij}}-\upsilon\left[T_i^{\gamma_{ij}-1}T_j^{\gamma_{ij}+1}+{\rm h.c.}\right]\right).
\end{equation}

\begin{figure}[htb]
  \centering
  \includegraphics[width=3.3in]{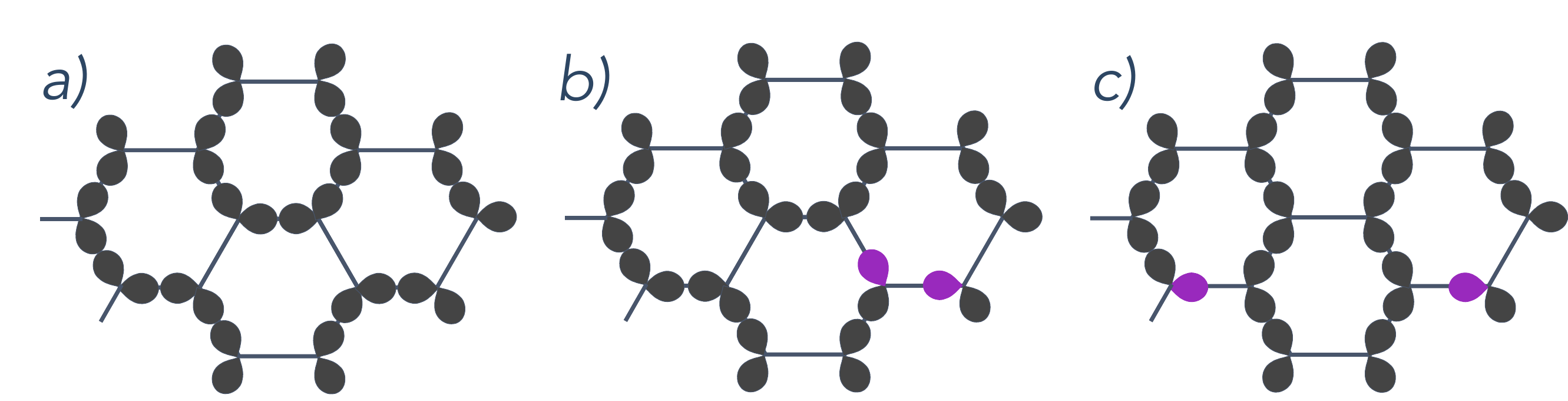}
  \caption{a) Orbital loop covering of the lattice. b,c) An elementary defect in the loop covering: b) at the loop
    cut, c) after part of the defect traveled. When the orbital loops
    are decorated by Haldane chains, the (purple) chain ends also
    carry a spin-$1/2$.}
  \label{fig:defects}
\end{figure}

{\em \textbf{Static loops.}---}To begin, we also focus on
$\upsilon=0$, in which case the Hamiltonian is exactly
soluble. Indeed, $H_{\rm orb}$ then reduces to a (classical) Potts model
$H_0=-\zeta\sum_{\langle
  ij\rangle}P_{i}^{\gamma_{ij}}P_{j}^{\gamma_{ij}}$ \cite{jackeli2008},
with $[P_i^\mu,P_j^\nu]=0$ and $[H_0,P_i^\mu]=0$ $\forall
i,j,\mu,\nu$. For $\zeta>0$, because $P_i^\gamma$ measures the
occupation of the $\gamma$ orbital at site $i$, on each bond $\langle ij\rangle$ the energy is minimized
when both orbitals $\gamma_{ij}$ at each end are filled, in which case
we say the bond is ``covered''. Because there
are two electrons per site, $\sum_{\gamma}P_i^\gamma=2$, the
configuration where two covered bonds stem out of every
site (forming a two-bond string) is favorable energetically (see Fig.~\ref{fig:defects}a)). The (fully-packed) loop coverings of the lattice implement
this condition throughout the lattice and constitute the
highly-degenerate ground state
manifold of $H_0$ \footnote{The loop covering manifold is isomorphic
  to the more familiar set of $1/3$-plateau states, or $2/3$-filling hard-core
  bosons, on the kagom\'{e} lattice by identifying a
  covered honeycomb {\em bond} with an ``up'' spin-$1/2$, or a hard-core boson, on the center of
  the bond.}. Owing to the orthogonality of $t_{2g}$
orbitals at the same site, all loop coverings are strictly orthogonal
to one another \cite{jackeli2007}. This is in contrast to many dimer models where the
dimers are two-spin-$1/2$ singlets. The
elementary excitations of $H_0$ (which take one out of the loop covering
manifold) are loop ``cuts:'' a loop is cut open, which creates nearby
two (``defect'') bonds covered by one orbital rather than two or zero (see
Fig.~\ref{fig:defects}b,c)). For $H_0$, while a loop cut costs an energy $\zeta$ and locally creates
two defect bonds, once created the two defect bonds may travel
infinitely apart at no further energy cost. This is of course reminiscent of the classical
spin ice problem, where a spin flip creates two monopoles which can (quasi-)freely
separate. 

\begin{figure}[htb]
  \centering
  \includegraphics[width=3.3in]{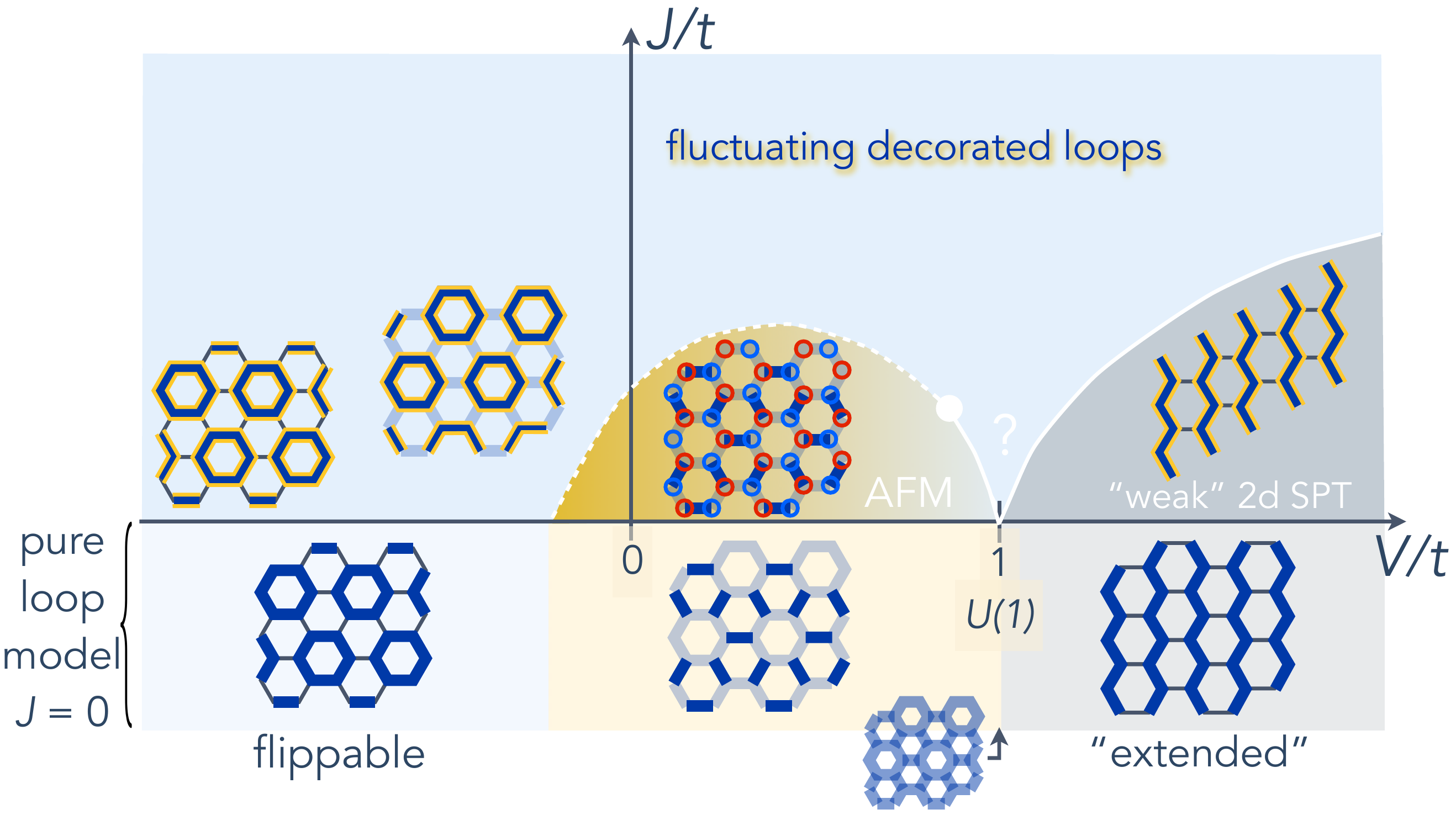}
  \caption{Phase diagram in 2d in the $V/t-J/t$ plane
    ($t=12\upsilon^3/\zeta^2$), for $J\geq0$ and $\zeta>0$. The phase
    diagram of the pure plain loop model (as obtained in Ref.~\onlinecite{schlittler2015a}) is shown below the
    horizontal axis. In the
    intermediate $J/t$ region, the location and nature of the phase
    transitions are speculative. 
    The solid and dashed lines and the
    white dot represent putative second and first order transitions,
    and critical end point, respectively. Thick blue lines represent
    orbital overlaps, yellow contours Haldane chains, and red and blue circles
    up and down spins.}
  \label{fig:2dphasediag}
\end{figure}

{\em \textbf{Orbital fluctuations.}---}We now consider a non-zero but small
$\upsilon\ll\zeta$. This gives dynamics to the loops,
since now $[H,P_i^\gamma]\neq0$. In degenerate perturbation theory in the
$\upsilon=0$ manifold, the
lowest-order effective Hamiltonian is
\begin{equation}
  \label{eq:5}
 H_{\rm eff}=-\frac{12\upsilon^3}{\zeta^2}\sum_{\hexagon} W_{\hexagon},
\end{equation}
where the sum is taken over all hexagons (or ``plaquettes'' of the
lattice), and corresponds to the ``flip'' terms given pictorially by:
\begin{equation}
  \label{eq:6}
  H_{\rm flip}=-t\sum_{\hexagon}\left(\left|\plaquettev\right\rangle\left\langle\plaquetteh\right|+\left|\plaquetteh\right\rangle\left\langle\plaquettev\right|\right)
\end{equation}
(microscopically, $t=12\upsilon^3/\zeta^2$). In Eq.~\eqref{eq:5}, in terms of
the $T$ operators
\begin{eqnarray}
  \label{eq:7}
  W_{\hexagon}&=&\mathcal{P}\,T^x_1T^y_2T^z_3T^x_4T^y_5T^z_6\,\mathcal{P}
\end{eqnarray}
where $\mathcal{P}$ is the projection onto the loop-covering
manifold and where the sites $1,..,6$ are defined around a hexagon as
in Fig.~\ref{fig:cubicenv}a). Hexagons with alternating covered and empty bonds, such as
those in Eq.~\eqref{eq:6} are called ``flippable.'' 

In dimer problems, it is customary to introduce a term, the
Rokhsar-Kivelson potential \cite{rokhsar1988}, which counts the number of
flippable plaquettes:
\begin{equation}
  \label{eq:8}
  H_{\rm
  RK}=V\sum_{\hexagon}\left(\left|\plaquettev\right\rangle\left\langle\plaquettev\right|+\left|\plaquetteh\right\rangle\left\langle\plaquetteh\right|\right),
\end{equation}
which can be written $H_{\rm
  RK}=V\sum_{\hexagon}$ $[\prod_{j=0}^2\mathsf{P}_{1+2j}^{\gamma_{1+2j,2+2j}}\mathsf{P}_{2+2j}^{\gamma_{1+2j,2+2j}}+\prod_{j=0}^2
\mathsf{P}_{2+2j}^{\gamma_{2+2j,3+2j}}\mathsf{P}_{3+2j}^{\gamma_{2+2j,3+2j}}]$,
where $\mathsf{P}^\mu_i=1-P^\mu_i$. $H_{\rm RK}$ is primarily used as a ``crutch'' to gain
insight from an accessible exactly soluble point.

\begin{figure}[htb]
  \centering
  \includegraphics[width=3.3in]{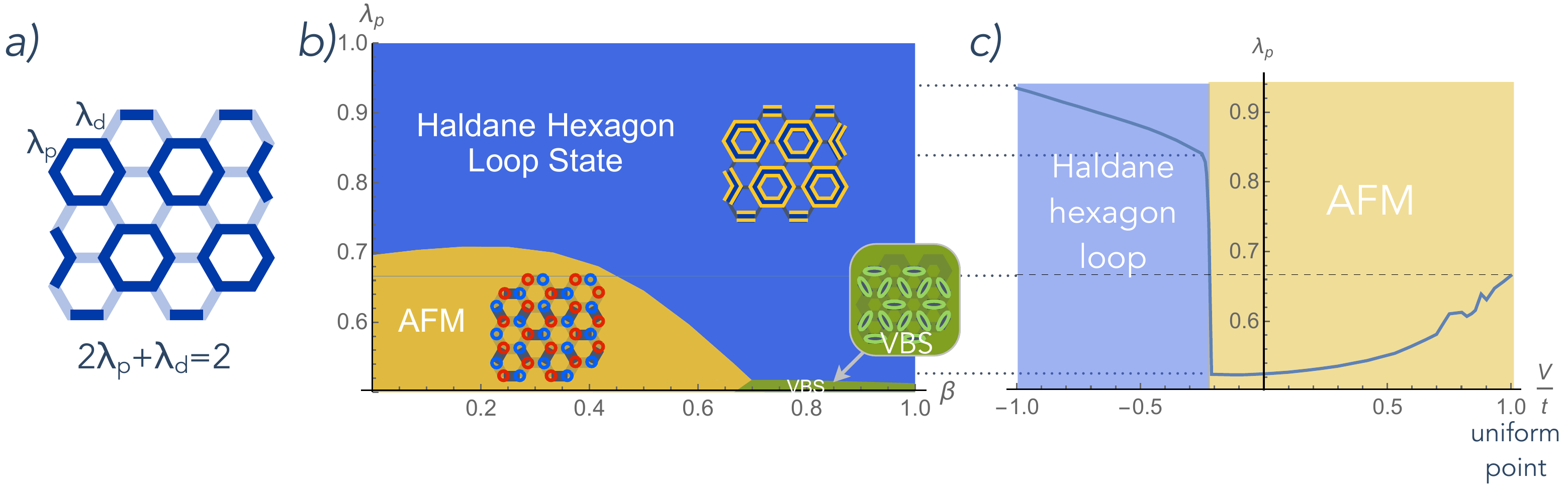}
  \caption{a) The non-uniform exchange pattern considered, with
    $\lambda_p,\lambda_d=\langle P_i^{\gamma_{ij}}P_j^{\gamma_{ij}}\rangle_{p,d}$ bonds forming plaquette and dimer
    structures, respectively. b) Phase
    diagram for the non-uniform bilinear-biquadratic model with the exchange
    pattern depicted in a), in the $\beta-\lambda_p$ plane, in
    a simple variational approach. c) $\lambda_p$ (shown in a)) as a function
    of $V/t$ as obtained from Ref.~\onlinecite{schlittler2015a}.}
  \label{fig:blbq}
\end{figure}

The loop model in general, and $\tilde{H}=H_{\rm flip}+H_{\rm
  RK}$ in particular, is in fact exactly dual to the dimer covering
model obtained from the loop one by ``swapping'' the covered and empty
bonds. The dimer model was studied in detail in several numerical
works \cite{moessner2001,schlittler2015a}, and the results adapted to
our loop model are presented below the horizontal axis in
Fig.~\ref{fig:2dphasediag} and in Fig.~\ref{fig:blbq}a,c), which we now
discuss. The phase
diagram of $\tilde{H}$ contains an exactly soluble point, that where
$V=t$, called the ``RK point,'' 
where the ground state is given by the
equal-weight quantum superposition of all loop coverings of the
lattice \cite{rokhsar1988}. This state, where the loops fluctuate wildly, has an emergent $U(1)$
(Coulombic) gauge field, and massive deconfined fractionalized
excitations, whose classical analogs are the
non-matching bonds obtained from loop cuts discussed above. It is a
$U(1)$ quantum orbital liquid, with a gapless (quadratic) photon mode. 

It is a well-known
result, however, that, in 2+1 dimensions, the deconfined phase of
$U(1)$ Coulombic gauge theories is unstable \cite{fradkin1979}, so that, in our model, the quantum
orbital liquid regime does not exist as a phase in an extended region
of the phase diagram, but survives only at the RK point. Away from the
RK point, the system instead releases its entanglement, breaks symmetries, and orders for
both $V>t$ and $V<t$ into the phases shown below the horizontal axis
in Fig.~\ref{fig:2dphasediag} and in Fig.~\ref{fig:blbq}a,c). At
$V/t>1$, the system immediately orders into ground states which feature static ``parallel'' chains
which extend through the whole system. For $V/t<1$, the system first
enters an ``intermediate phase'' where $0<\left\langle
P_{i}^{\gamma_{ij}}P_j^{\gamma_{ij}}\right\rangle<1$, before hitting a
first-order phase transition below which the system favors one of the
three ``maximally flippable'' ``hexagon loop crystal''
configurations (see figure in Supp.\ Mat.). 

We note that the model presented here
can be generalized to three dimensions (on the hyperhoneycomb
lattice \cite{takayama2015}, which shares with the honeycomb lattice the same essential
ingredients), where Coulombic phases of $U(1)$ gauge theories are
stable \cite{fradkin1979,banerjee2008,savary2012coulombic}. Details are beyond the
scope of this paper, but will be addressed in an upcoming
publication \cite{savary2015}. Stable two-dimensional generalizations, such as
those allowing for a $\mathbb{Z}_2$ spin liquid, are also possible.

{\em \underline{\textbf{Spins.}}---}We now finally introduce the spins,
i.e.\ consider $J\neq0$. First, we note that the spin operators appear
only in
\begin{equation}
  \label{eq:3}
  H_J=\sum_{\langle ij\rangle}\tilde{J}_{ij}\left(\mathbf{S}_i\cdot\mathbf{S}_j+\beta(\mathbf{S}_i\cdot\mathbf{S}_j)^2\right),
\end{equation}
where $\tilde{J}_{ij}=J P_{i}^{\gamma_{ij}}P_j^{\gamma_{ij}}$, which,
when considered as a 1d problem with constant $\tilde{J}_{ij}>0$ and
$-1\leq\beta\leq1$, realizes the Haldane phase (and in particular the AKLT state described in the introduction
at $\beta=1/3$). Notably, the spin exchange is ``modulated'' by the
operator $P_i^{\gamma_{ij}}P_j^{\gamma_{ij}}$, and vanishes when
$P_i^{\gamma_{ij}}P_j^{\gamma_{ij}}=0$. Therefore, when the system
forms orbital loops, the problem in spin space reduces to a collection of purely one-dimensional periodic $S=1$ Hamiltonians,
which are minimized by entering the Haldane phase. This leads to the appearance of new
structures, namely Haldane decorated loops, where each orbital
loop subtends a Haldane chain, 
$|\hat{\mathcal{L}}\rangle=|\mathcal{L}\rangle\otimes|\psi_{\rm Haldane}\rangle$,
where $|\mathcal{L}\rangle$ is an orbital loop, and $|\psi_{\rm
  Haldane}\rangle$ the Haldane spin ground state. Interestingly, the
decoration in general 
introduces a length-dependent energy density. Indeed, away from $\beta=1/3$, where
the energy density (the energy per site, or bond) of periodic AKLT
chains is independent of their length, the energy density of
length-six loops is always smaller than that of longer loops (see
Supp.\ Mat.\ for results obtained using DMRG). In turn, this has
consequences on the energetics of the loop coverings, which may become inequivalent.


{\em \textbf{Static coverings: $\upsilon=0$.}---}In the absence of orbital fluctuations,
i.e.\ when $\upsilon=0$, the ground state manifold of the pure orbital model is
that of all loop-coverings, as discussed above. In particular, all
loop coverings are degenerate in energy, regardless of the
distribution of their loop lengths. If we now consider $J>0$ and $J\ll\zeta$,
at first order in perturbation theory in $\zeta/J$ ($H_J$ perturbs $H_0$), 
spin states 
break the degeneracy of the loop
coverings, following
$\langle\hat{\mathcal{C}}|H_{J}|\hat{\mathcal{C}}\rangle$ where the
$|\hat{\mathcal{C}}\rangle$ are the otherwise-degenerate decorated
loop coverings. 
At $\beta=1/3$, we retain
an exact degeneracy between Haldane-decorated loop configurations,
which all together form the ground state manifold, while, away from
$\beta=1/3$, $\langle\hat{\mathcal{C}}|H_{J}|\hat{\mathcal{C}}\rangle$
is only minimized when the
system forms one of three equivalent ``hexagon crystal'' states where
the lattice is covered by decorated loops of length six (see Supp.\ Mat.). 
This static-orbital regime corresponds to
the infinite $J/t$ limit on Fig.~\ref{fig:2dphasediag}. 

We now introduce the orbital kinetic terms, distinguishing between
different $J/\upsilon$ regimes. 

\begin{figure}[htb]
  \centering
  \includegraphics[width=3.3in]{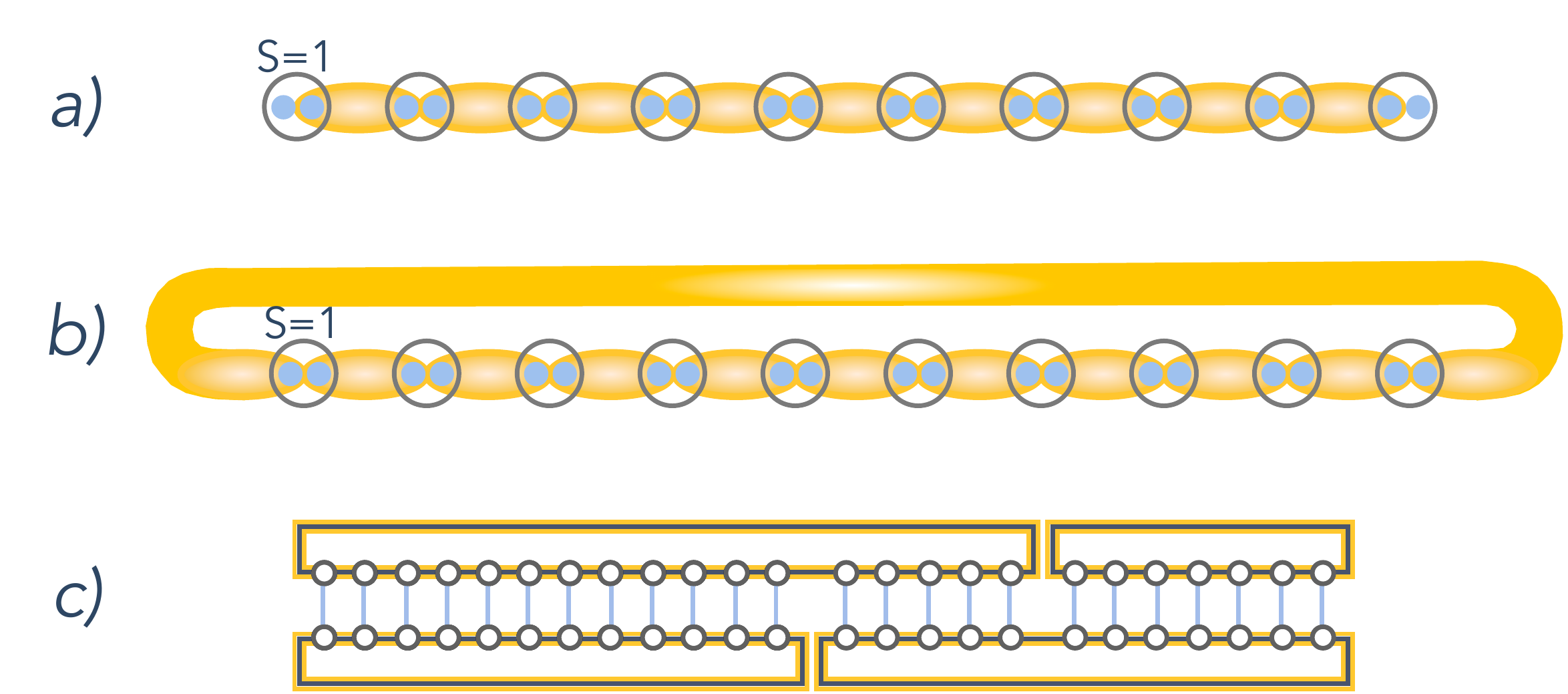}
  \caption{a) Open and b) closed (periodic) AKLT chains, 
    c) MPS representation of the transfer matrices
    for the overlap between different AKLT chains 
    ``coverings.'' 
  }
  \label{fig:mps}
\end{figure}

{\em \textbf{Large $J/\upsilon$ limit.}---}In the large $J/t$ limit,
we first consider the Hamiltonian 
\begin{equation}
  \label{eq:9}
    H_{\rm stat}=\sum_{\langle
    ij\rangle}P_{i}^{\gamma_{ij}}P_{j}^{\gamma_{ij}}\left(-\zeta+J\left(\mathbf{S}_i\cdot\mathbf{S}_j+\beta(\mathbf{S}_i\cdot\mathbf{S}_j)^2\right)\right),
\end{equation}
and introduce the kinetic terms $\upsilon$ in perturbation
theory. Even with $J>0$ (and possibly $J\sim\zeta$), the eigenstates of $H_{\rm stat}$ are still eigenstates of
the $P_i^\mu=\hat{n}_i^\mu$, and as discussed above $\tilde{J}_{ij}=JP_{i}^{\gamma_{ij}}P_{j}^{\gamma_{ij}}$ is
only nonzero when
$\hat{n}_i^{\gamma_{ij}}=\hat{n}_j^{\gamma_{ij}}=1$. Therefore 
ground states of $H_{\rm stat}$ belong to the set of decorated loop
covering, so long as $-\zeta+\epsilon_{\rm Hald\,cov}<0$, where
$\epsilon_{\rm Hald\,cov}$ is the energy density of the collection of all pure Haldane chains
in the covering. For example, for $\beta=0$ (resp.\ $\beta=1/3$), this is true for any
$\zeta>\epsilon_{L=\infty}(0)\approx-1.40J$ (resp.\ $\zeta>\epsilon_{L=\infty}(1/3)=-2/3J$). 

When $\beta=1/3$, the manifold of
decorated loop coverings (which are ground states of $H_{\rm stat}$) is highly degenerate, and we call $\mathfrak{P}$ the projector
onto this manifold. 
In perturbation theory in small
$\upsilon/(-\zeta+\epsilon_{\rm Hald\,cov})$, our analysis
of the pure orbital model informs us that the lowest order
contribution arises at third order, provided $H_{\rm eff}'=\mathfrak{P}(H_{\rm
  flip}\otimes\mathds{1}_\mathbf{S})\mathfrak{P}$ does not identically
vanish. Indeed, $H_{\rm flip}$ is the lowest-order orbital-space
Hamiltonian to take one orbital loop covering into another, and one
must check that the corresponding Haldane loop coverings have
non-zero overlap. Remarkably, we find that the overlap between two
AKLT loop coverings (``cov'') which differ by a single ``plaquette flip'' is always
equal to $1/4$, up to exponentially-small corrections in the lengths
of the (rearranged) loops, i.e.\
\begin{equation}
  \label{eq:10}
  \langle\hat{\mathcal{C}}_1|(H_{\rm
  flip}\otimes\mathds{1}_\mathbf{S})|\hat{\mathcal{C}}_2\rangle =\langle{\rm AKLT\,cov}_1| {\rm AKLT\,cov}_2\rangle\approx\frac{1}{4},
\end{equation}
regardless of how many loops are connected by a plaquette flip
\footnote{Also note that on a bipartite lattice, like the honeycomb,
  the sign of singlets across bonds can be unambiguously
  systematically chosen, in contrast to e.g.\ in Ref.~\onlinecite{wang2015}.}. This
result is obtained using the matrix product state (MPS) formalism (see Supplemental Material). In fact,
Eq.~\eqref{eq:10} is a special case of
$\langle\otimes{\rm AKLT}_1|\otimes{\rm AKLT}_2\rangle\approx1/2^{n_{\rm cuts}-1}$, where $n_{\rm
  cuts}$ is the number of loop cuts needed to connect $|\otimes{\rm
  AKLT}_{1,2}\rangle$ (see Fig.~\ref{fig:mps}d)). This momentous result is a consequence of the ultra
short range entanglement of the AKLT state, which is ``close''
to being a product of single-{\em site} states. 

Away from $\beta=1/3$, but within the Haldane phase ($|\beta|\leq1$),
we expect the same results to hold since entanglement properties are
characteristic of a phase. Therefore, at large $J/t$, and for
$\beta=1/3$ and $V/t=1/4$ (``decorated RK point''), the
system is ``close to'' \footnote{The overlap Eq.~\eqref{eq:10} is only
approximately equal to $1/4$, and depends---albeit exponentially---on
loop length; this could in fact bear additional interesting consequences \cite{penc1995}.} a $U(1)$ phase with spinful
fractional excitations, in the sense that it contains large
fluctuating Haldane-decorated loops. This is a state the model
was designed to achieve. The lowest-energy excitations are either loop
cuts (necessarily accompanied by a Haldane chain cut) or pure Haldane
chain excitations, depending on the distribution of loop length and
values of $\zeta$ and $\beta$ \footnote{For example, for $L=6$, at $\beta=0$ and
$1/3$, $\epsilon_{6}^1-\epsilon_{6}^0\approx0.120J$ and
$\epsilon_{6}^1-\epsilon_{6}^0\approx0.116J$, respectively, and
the energy density differences
between an open and a closed chain of length six are
$\breve{\epsilon}_{6}^0-\epsilon_{6}^0\approx0.208J$ and
$\breve{\epsilon}_{6}^0-\epsilon_{6}^0=J/9$, respectively.}. Local loop cuts generate two orbital chain ends, which
are decorated by spin-1/2. In three dimensions, where the orbital
$U(1)$ deconfined phase survives away from the RK point, these chain
ends are deconfined spinons.


{\em \textbf{Small $J/\upsilon$ limit.}---}In the small $J/t$ limit, the orbital-only
model, i.e.\ $H_{\rm orb}$ from Eq.~\eqref{eq:3}, is solved first and
spin exchange ($J$) is then introduced perturbatively. This 
results in an
effective exchange pattern for the spins. More precisely, when
$J=0$, spin space is completely degenerate, but the orbital ground
state is a priori unique (or discretely degenerate due to
symmetry-related states). Therefore, upon introducing $J$, in
degenerate perturbation theory, we have, at first order,
\begin{eqnarray}
  \label{eq:11}
  H_{\rm
  eff}''&=&\sum_{\langle ij\rangle}\mathfrak{p}\left[JP_i^{\gamma_{ij}}P_j^{\gamma_{ij}}\left(\mathbf{S}_i\cdot\mathbf{S}_j+\beta(\mathbf{S}_i\cdot\mathbf{S}_j)^2\right)\right]\mathfrak{p}\\
&=&\sum_{\langle ij\rangle}\left\langle P_i^{\gamma_{ij}}P_j^{\gamma_{ij}}\right\rangle J\left(\mathbf{S}_i\cdot\mathbf{S}_j+\beta(\mathbf{S}_i\cdot\mathbf{S}_j)^2\right),
\end{eqnarray}
where $\mathfrak{p}$ is the projector onto the $J=0$ ground state, and
$\left\langle.. \right\rangle$ is the expectation value taken in this ground
state. We obtain the phase diagram within the pure loop model by using the results from
Ref.~\onlinecite{schlittler2015a} and a simple variational approach
(see Supp.\ Mat.\ for a detailed derivation). Notably, the
Haldane loop decoration in the extended phase at large $V/t>0$
gives way to a two-dimensional weak SPT phase, with neutral Kramers
doublet \footnote{Time-reversal symmetry is also preserved in that phase.} edge
states (the edge orbitals decorated by spin-$1/2$ degrees of freedom) protected by translational symmetry, provided the boundaries
are appropriately chosen (and, strictly speaking, provided a weak
coupling at the boundary exists).  It is noteworthy that this phase is
realized spontaneously, 
i.e.\ this is {\em not} an explicit ``chain-stacking'' construction. To our
knowledge, this is the first such example in the literature.


In this regime ($\zeta>0$, $\zeta\gg J>\upsilon$) low-energy excitations
are expected to occur in the spin sector. In the antiferromagnetic (AFM) phase, those are simply the
conventional spin flips. In the decorated chain phases, the elementary
excitations are those of the (gapped) Haldane chains. Remarkably, in
the length-six loop state, which is a product state of decorated
hexagon loops, the excitations are local, but a weak coupling between
the hexagons (e.g.\ when $\upsilon\neq0$) will lead to
slightly-dispersive ``Haldane gap waves,'' observable for example in
neutron scattering. 


The
results derived above are summarized in the phase diagram in Fig.~\ref{fig:2dphasediag}. 

{\em \underline{\textbf{Discussion.}}---}In summary, we have exposed a
physical mechanism for the realization
of fluctuating Haldane chains in spin-orbital models in two
dimensions. To do so, we presented a realistic and analytically-tractable spin-orbital Hamiltonian on the honeycomb
lattice, with a rich phase diagram, featuring exotic
phases built out of Haldane chains. Among those are a translational-symmetry protected topological phase, with spin-1/2 edge excitations,
a Haldane hexagon loop ``crystal,'' with ``Haldane gap wave'' excitations, and a regime with fluctuating Haldane
chains coupled to underlying ``orbital loops.'' On the
three-dimensional hyperhoneycomb lattice the latter becomes a Coulombic quantum
spin-orbital liquid, a unique example in the spin-orbital literature of a
controllable model where both the spin and orbital sectors are
``disordered.'' Moreover, supplementing the model with additional terms is likely
to allow accessing more phases and possibly interesting phase
transitions. In fact, many more avenues---in several different
fields---will be worth exploring further. For example, the
quantum spin liquid 
can be induced not only by taking the model to three dimensions but also by turning it into a
$\mathbb{Z}_2$ liquid. The variation of the parameter $\zeta$ or the number of
electrons per site may also lead to interesting problems and phase
transitions. In general, the
highlighted mechanism 
will hopefully be an important stepping stone for future studies to realize Haldane chains
and other low-dimensional structures in higher dimensions.

Most exciting would certainly be the discovery in real materials of some of the
phenomena described here. This model is relevant to insulating
honeycomb materials with two electrons in degenerate $t_{2g}$ orbitals
and
large Hund's coupling to enforce $S=1$. In practice, materials need to
have (at least approximate) cubic symmetry, weak spin-orbit
coupling and large direct orbital overlap. Therefore, materials based on Ru, Ni, V, etc.\
na\"{i}vely appear as potential candidates. Regardless, it will be
important and interesting to study the breaking
of any of these constraints, through e.g.\ spin-orbit coupling or symmetry lowering, inevitable at some level in real materials. Magneto-elastic
coupling should also be investigated. It might well play a role similar to the $V$ term in
stabilizing the ``extended'' or ``flippable'' phases.



\acknowledgements{It is a pleasure to thank George Jackeli, Giniyat
Khaliullin, Adam Nahum, and especially Leon Balents for many
enlightening discussions and comments, as well as Max Metlitski and
Xiao-Gang Wen for useful discussions and T.\ Senthil for bringing to
our attention the problem of the realization of Haldane chains in
$S=1$ materials. This work was generously supported by a postdoctoral
fellowship from the Gordon and Betty Moore Foundation through the
EPiQS initiative, grant number GBMF4303. It was finalized
while at KITP and partially supported by grants NSF-DMR-11-21053 and
NSF-PHY-11-25915. DMRG calculations were performed using the itensor package.}

\bibliography{honeycomb-bib.bib}

\appendix

\section{Effective orbital operators}
\label{sec:effective-operators}

\subsection{Construction of the states and operators}
\label{sec:constr-stat-oper}

Let us consider two electrons per site, and degenerate $t_{2g}$
($d_{xy}$, $d_{xz}$ and $d_{yz}$) orbitals at each site, and a very large intra-orbital $U$, so that
there is only one electron per orbital, a large Hund's coupling
$J_{\rm H}$ so that $S=1$, and no spin orbit coupling. We define the
states of the three-dimensional orbital space to be $|x\rangle$,
$|y\rangle$ and $|z\rangle$, such that, if $|0\rangle$ is the Fock
space vacuum at a given site for spinless electrons and
$c^\dagger_{\mu\nu}$ creates a spinless electron in orbital $d_{\mu\nu}$:
\begin{equation}
  |\gamma\rangle=c_{\gamma+1,\gamma+2}(c^\dagger_{yz}c^\dagger_{xz}c^\dagger_{xy}|0\rangle).
\end{equation}
The normalization is chosen such that
$\langle\gamma|\gamma'\rangle=\delta_{\gamma,\gamma'}$. Note that an
unimportant (convention-dependent) choice of phase was made.

We now define the operators $L^c$, $c=x,y,z$ according to:
\begin{equation}
  L^c=\frac{-i}{2}\sum_{a,b}\epsilon_{abc}(|a\rangle\langle
  b|-|b\rangle\langle a|),
\end{equation}
which may be rewritten:
\begin{equation}
  L^\gamma=-i\left[|\gamma+1\rangle\langle\gamma-1|-|\gamma-1\rangle\langle\gamma+1|\right].
\end{equation}
These operators obey $L^\gamma|\gamma\rangle=0$, and $\mathbf{L}^2=2$. One may check that these operators obey the angular momentum algebra
commutation relations. To form a complete basis of Hermitian operators acting in
our three-dimensional space, we need six more Hermitian operators,
which we choose to be $(L^a)^2=P^a$ and $\{L^a,L^b\}=T^c$, $a\neq b$. Note
that:
\begin{equation}
  P^\gamma=(L^\gamma)^2=|\gamma+1\rangle\langle\gamma+1|+|\gamma-1\rangle\langle\gamma-1|=1-|\gamma\rangle\langle\gamma|
\end{equation}
and
\begin{equation}
  T^\gamma=\{L^{\gamma+1},L^{\gamma-1}\}=-\left(|\gamma+1\rangle\langle\gamma-1|+|\gamma-1\rangle\langle\gamma+1|\right).
\end{equation}

\subsection{Effective operators as ``rotation'' and projection operators}
\label{sec:effect-oper-as}

$L^\gamma|\gamma\rangle=0$ so the projection operator onto the
$|\gamma\rangle$ component is $\mathsf{P}_\gamma=1-(L^\gamma)^2$. The
projection operator in Eq.~\eqref{eq:2} is
$P_\gamma=1-\mathsf{P}_\gamma=(L^\gamma)^2$. In particular:
\begin{equation}
  \label{eq:17}
  P_\gamma|\gamma\rangle=0,\qquad P_\gamma|\gamma\pm1\rangle=|\gamma\pm1\rangle.
\end{equation}


Disregarding phase factors:
\begin{equation}
  \label{eq:15}
  \begin{cases}
    L^\gamma|\gamma\rangle=0\\
T^\gamma|\gamma\rangle=0
  \end{cases},\qquad
  \begin{cases}
    L^\gamma|\gamma\pm1\rangle\propto|\gamma\mp1\rangle\\
T^\gamma|\gamma\pm1\rangle\propto|\gamma\mp1\rangle
  \end{cases}.
\end{equation}

\subsection{Spin-orbital model}
\label{sec:spin-orbital-model}

In the orbital sector, the coupling Hamiltonian between two sites $1$
and $2$ connected by an $x$-type bond, as defined above, takes the
form:
\begin{widetext}
  \begin{eqnarray}
      H_{12}^{orb}&=&
  \begin{pmatrix}
    {L_{1}^x}^2&{L_1^y}^2&{L_1^z}^2
  \end{pmatrix}
\begin{pmatrix}
   \mathcal{J}_1&\mathcal{J}_4&\mathcal{J}_4\\
    \mathcal{J}_4&\mathcal{J}_2&\mathcal{J}_3\\
    \mathcal{J}_4&\mathcal{J}_3&\mathcal{J}_2
  \end{pmatrix}
\begin{pmatrix}
  {L_{2}^x}^2\\{L_2^y}^2\\{L_2^z}^2
\end{pmatrix}
+
  \begin{pmatrix}
    L_{1}^x&L_1^y&L_1^z
  \end{pmatrix}
\begin{pmatrix}
    \mathcal{J}_5&\mathcal{J}_8&-\mathcal{J}_8\\
    \mathcal{J}_8&\mathcal{J}_6&\mathcal{J}_7\\
    -\mathcal{J}_8&\mathcal{J}_7&\mathcal{J}_6
  \end{pmatrix}
\begin{pmatrix}
  L_{2}^x\\L_2^y\\L_2^z
\end{pmatrix}\nonumber
\\
&&+  \begin{pmatrix}
    \{L_{1}^y,L_1^z\}&\{L_{1}^x,L_1^z\}&\{L_{1}^x,L_1^y\}
  \end{pmatrix}
\begin{pmatrix}
   \mathcal{J}_9&\mathcal{J}_{12}&-\mathcal{J}_{12}\\
    \mathcal{J}_{12}&\mathcal{J}_{10}&\mathcal{J}_{11}\\
   - \mathcal{J}_{12}&\mathcal{J}_{11}&\mathcal{J}_{10}
  \end{pmatrix}
\begin{pmatrix}
  \{L_{2}^y,L_2^z\}\\\{L_{2}^x,L_2^z\}\\\{L_{2}^x,L_2^y\}
\end{pmatrix}.
    \label{eq:21}
\end{eqnarray}
\end{widetext}
In
Eq.~\eqref{eq:1}, if $J=0$, $\mathcal{J}_1=-\zeta$, $\mathcal{J}_7=-\upsilon_1$, 
$\mathcal{J}_{11}=-\upsilon_2$, and
all others zero. While the number of parameters is large (12), many of
them are expected to be zero, physically. For example, it is unclear
whether it is possible to obtain terms which
involve a single power of angular momentum at each site from standard
superexchange calculations \cite{khaliullin2005,khaliullin2013}. 

Upon introducing the spin degrees of freedom, in principle, each
independent coefficient may be a spin Hamiltonian of the form
(for no spin-orbit coupling):
\begin{equation}
  \label{eq:22}
  \mathcal{J}_p=A_p+B_p\mathbf{S}_1\cdot\mathbf{S}_2+C_p(\mathbf{S}_1\cdot\mathbf{S}_2)^2,
\end{equation}
where $p=1,..,12$ labels the independent terms. In Eq.~\eqref{eq:1}, we
took $\mathcal{J}_1=-\zeta+J(\mathbf{S}_1\cdot\mathbf{S}_2+\beta(\mathbf{S}_1\cdot\mathbf{S}_2)^2)$, i.e.\ $A_1=-\zeta$, $B_1=J$,
$C_1=J\beta$, and $\mathcal{J}_7=-\upsilon_1$, i.e.\ $A_7=-\upsilon_1$
and $B_7=C_7=0$, and $\mathcal{J}_{11}=-\upsilon_2$, i.e.\ $A_{11}=-\upsilon_2$
and $B_{11}=C_{11}=0$ and all other terms zero.

\section{Details of perturbation theory}
\label{sec:deta-pert-theory}

Here we focus on orbital space, i.e.\ set $J=0$, and give a few details
for the degenerate perturbation theory in
\begin{equation}
  \label{eq:42}
  H_{\rm kin}=-\upsilon\sum_{\langle ij\rangle}(T_i^{\gamma_{ij}-1}T_j^{\gamma_{ij}+1}+T_i^{\gamma_{ij}+1}T_j^{\gamma_{ij}-1})
\end{equation}
onto the manifold of loop
coverings of the lattice, valid when $\zeta>0$ and $\zeta\gg\upsilon$. The effective Hamiltonian
is
\begin{equation}
  \label{eq:41}
  H_{\rm eff}=\mathcal{P}H_{\rm kin}\frac{1-\mathcal{P}}{H-E_0}H_{\rm kin}\frac{1-\mathcal{P}}{H-E_0}\cdots \frac{1-\mathcal{P}}{H-E_0}H_{\rm kin}\mathcal{P},
\end{equation}
where $H_{\rm kin}$ appears as many times as the order in perturbation
theory. 

\begin{figure}[htb]
  \centering
  \includegraphics[width=3.3in]{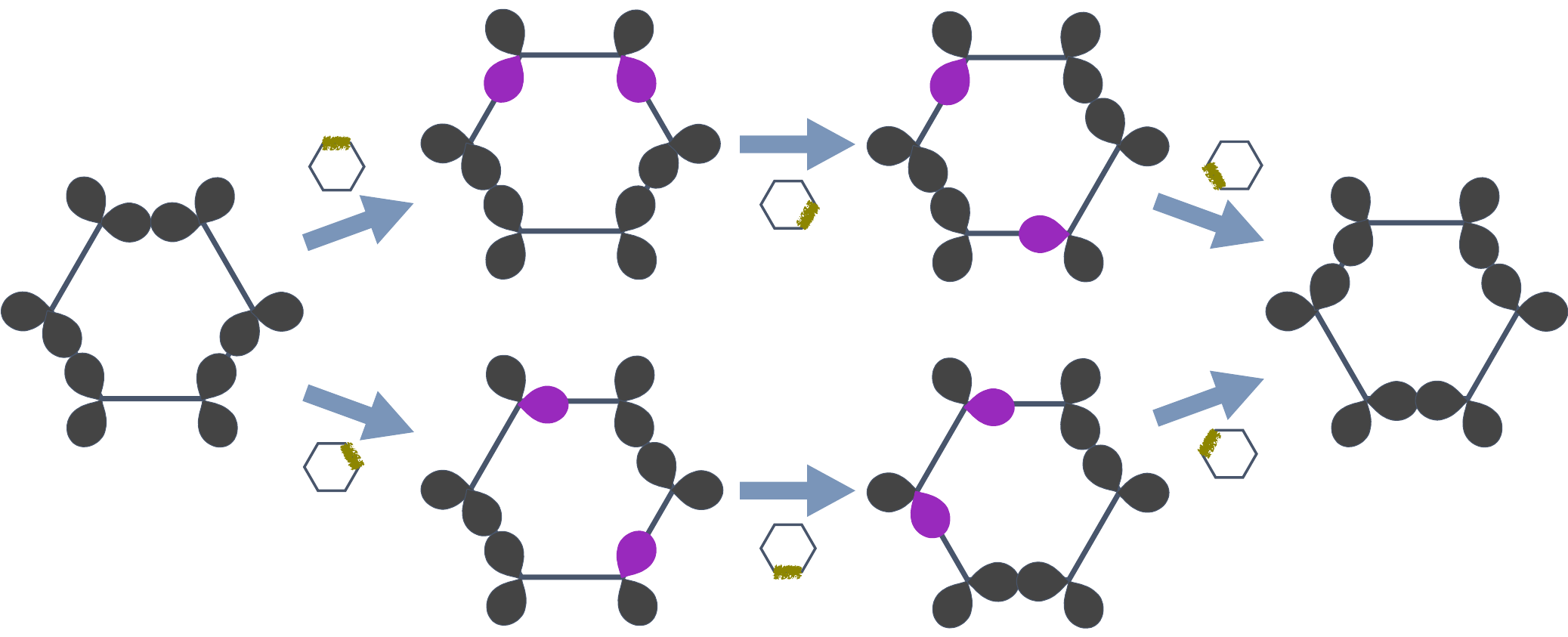}
  \caption{Flippable plaquette adventure through third-order
    perturbation theory. The small hexagons next to the arrows show
    the bond on which $H_{\rm kin}$ is applied at that order.}
  \label{fig:PT}
\end{figure}

Consider a ``flippable'' plaquette. Acting once with $H_{\rm kin}$ on
any bond which belongs to the plaquette creates two ``defect''
bonds (this configuration does not belong to the loop covering
manifold), with the new plaquette state looking like on Fig.~\ref{fig:PT}. The
energy of this configuration is that of a loop cut, i.e.\
$\zeta$. Acting a second time with $H_{\rm kin}$, with the ``active'' bond
operator one bond away from the first active bond creates another
configuration of energy $\zeta$. Only at third order is the system
brought back to the loop manifold. There are twelve ($=6\times2$) ways
to achieve this. It is noteworthy that including many other terms from
Eq.~\eqref{eq:21} will not produce a lower-order contribution.

\section{Haldane chain energy}
\label{sec:haldane-chain-energy}

In this appendix we investigate the energy density (energy divided by
the number of sites) of $S=1$ loops in the Haldane phase as a function
of their length. 

\subsection{AKLT chains}
\label{sec:aklt-chains}

At the AKLT point, the energy density is independent
of the loop length. Indeed the AKLT Hamiltonian may be rewritten as
\begin{equation}
  H_{\rm
    AKLT}=\frac{1}{4}\sum_i\left[(\mathbf{S}_i+\mathbf{S}_{i+1})^2\left((\mathbf{S}_i+\mathbf{S}_{i+1})^2-2\right)+{\rm
    const}\right],
\end{equation}
i.e.\ as the sum of the projectors (with equal {\em positive}
coefficient) onto the $S^{\rm tot}=2$ sector (i.e.\ $(\mathbf{S}^{\rm tot})^2=2(2+1)=6$) of the $\mathbf{S}_i^{\rm
  tot}=\mathbf{S}_i+\mathbf{S}_{i+1}$ operator. This means that the
ground state will have zero components in the $S=2$ sector. Then, the
energy is independent of chain length. Hence, at first order in
perturbation theory, the spins do not lift the degeneracy of the loop
coverings at the spin AKLT point.


\subsection{Numerical results away from the AKLT point}
\label{sec:numerical-results}

\begin{figure}[htb]
  \centering
  \includegraphics[width=3.3in]{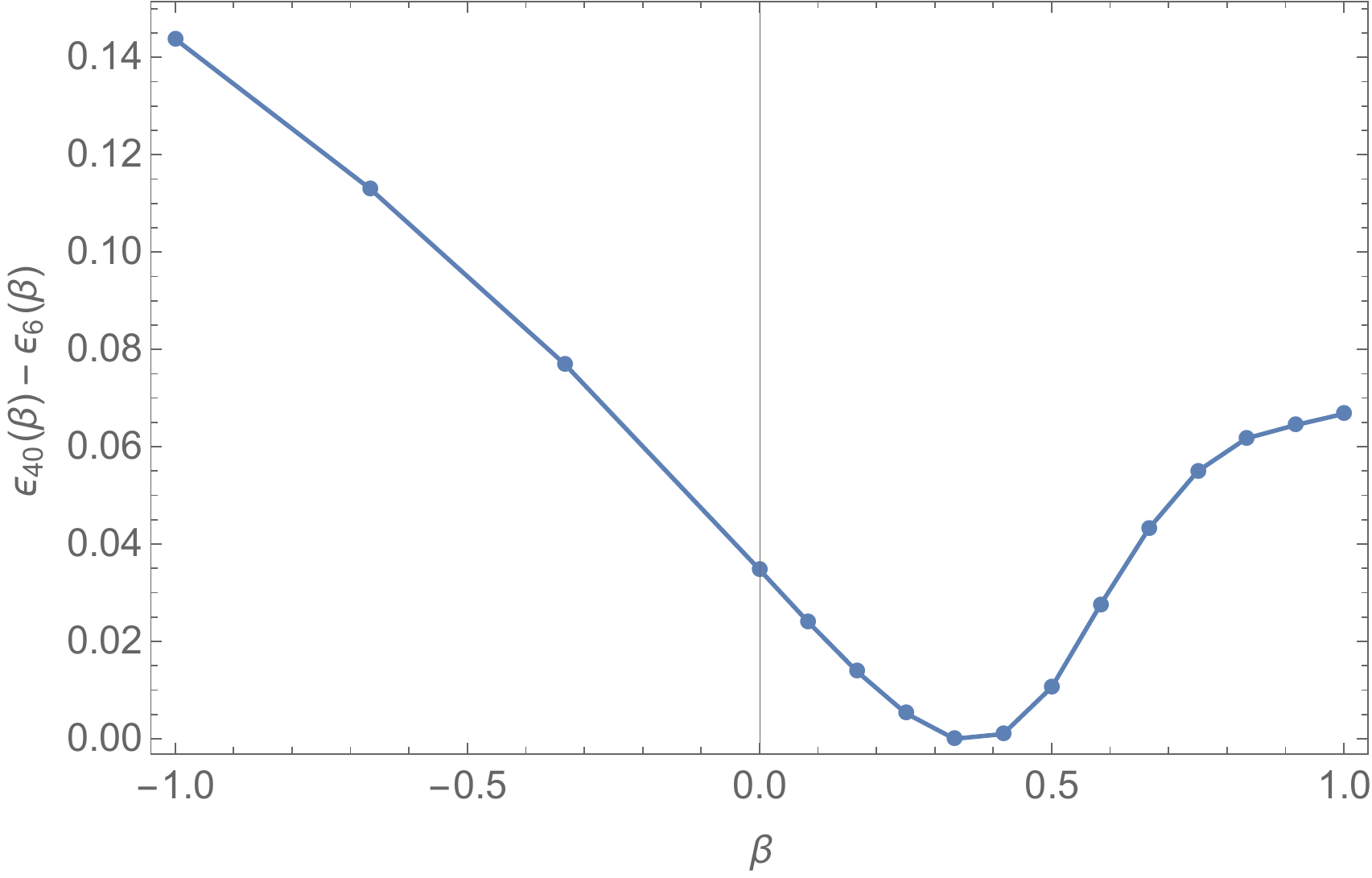}
  \caption{Difference in ground state energy density of periodic
    chains of length 40 and 6, as a function of $\beta$, as calculated
    in DMRG, in units of $J$. 
    The energy density at length 40 is ``assumed'' to be close to that of infinite-length chains.}
  \label{fig:length640}
\end{figure}

We performed exact diagonalization for the Hamiltonian in
Eq.~\eqref{eq:1} on chains with periodic boundary conditions for up to
length 7. The results seem to indicate that, away from the AKLT point
$\beta=1/3$, the energy density of {\em closed} even-length loops (relevant for the
honeycomb and hyperhoneycomb lattices) increases with loop
length. Results obtained in DMRG for longer closed loops with the use of the itensor package
confirm that the energy density of loops of length 40 is always larger
than that of length 6, see Fig.~\ref{fig:length640} (and
Ref.~\onlinecite{sun1992numerical} for Monte Carlo results at $\beta=0$).

\section{Haldane covering overlaps}
\label{sec:hald-cover-overl}

\subsection{At the AKLT point: MPS formalism}
\label{sec:at-aklt-point}

\begin{figure}[htb]
  \centering
  \includegraphics[width=3.3in]{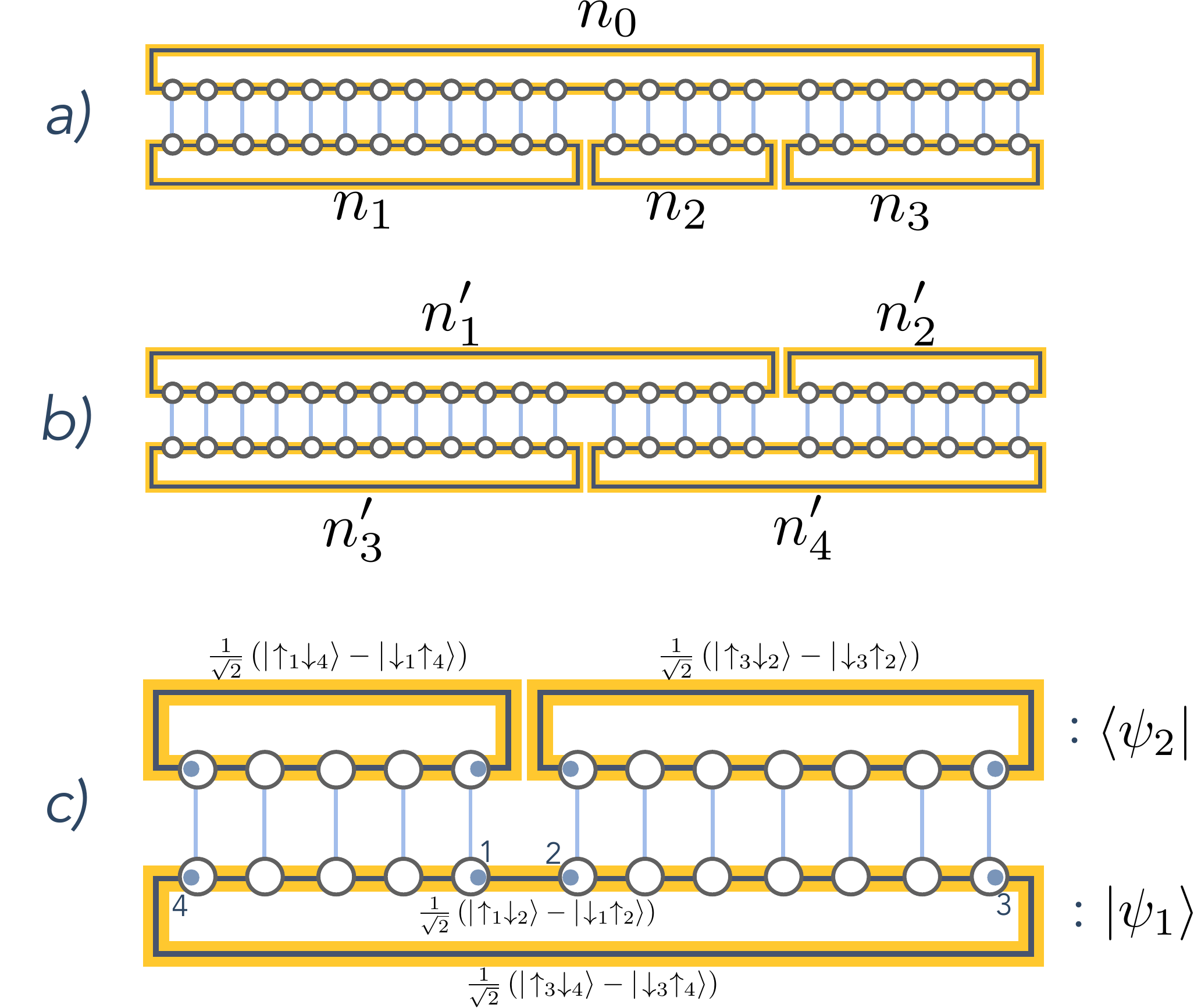}
  \caption{Transfer matrix overlap representation in the matrix
    product state formalism. a) Overlap between three loops and one
    loop, b) overlap between two sets of two loops, c) overlap between
  one loop and two loops, emphasizing the role of the fractional
  degrees of freedom near the cuts.}
  \label{fig:mpssuppmat}
\end{figure}

The AKLT loop covering overlaps are calculated with the matrix product
state formalism. An exact representation of the AKLT wavefunction is
given as an MPS:
\begin{equation}
  \label{eq:14}
  |\psi\rangle_{\rm AKLT}=\sum_{\{\sigma_i=0,\pm1\}}{\rm Tr}\left[\mathsf{M}(\sigma_1)\cdots\mathsf{M}(\sigma_N)\right]|\sigma_1\cdots\sigma_N\rangle,
\end{equation}
when the chain is a closed loop of length $N$ (we use the notations from Ref.~\onlinecite{balents2014}) with the following matrices $\mathsf{M}$:
\begin{equation}
  \label{eq:26}
\begin{cases}
  \mathsf{M}(\sigma=0)=-\sqrt{\frac{1}{3}}\sigma^z\\
  \mathsf{M}(\sigma=\pm1)=-\sqrt{\frac{2}{3}}\sigma^\pm
\end{cases},
\end{equation}
with $\sigma^\mu$ the Pauli matrices, and with the norm $|\psi|^2$:
\begin{equation}
  \label{eq:27}
  \langle\psi|\psi\rangle={\rm
    Tr}\mathsf{T}^N,\qquad\mbox{where}\qquad
\mathsf{T}=\sum_{\sigma=\pm1,0}\mathsf{M}^*(\sigma)\otimes\mathsf{M}(\sigma).
\end{equation}
We then compute the overlap between different types of coverings
connected by a single plaquette flip. The difference between those
configurations lies purely in the number and lengths of the loops
``touching'' the flippable plaquette of interest, before and after the
plaquette flip. Within the MPS formalism, these overlaps are given by:
\begin{itemize}
\item three loops connected to one:
\begin{eqnarray}
  \label{eq:31}
&&\langle{\rm AKLT}_0|{\rm AKLT}_1,{\rm AKLT}_2,{\rm
   AKLT}_3\rangle\\
&&\qquad\qquad\qquad\qquad\qquad\qquad=\frac{{\rm
    Tr}[\mathsf{T}_1^{n_1}\mathsf{T}_2^{n_2}\mathsf{T}_3^{n_3}]}{\sqrt{{\rm Tr}[\mathsf{T}^{n_0}]{\rm
    Tr}[\tilde{\mathsf{T}}_1^{n_1}\tilde{\mathsf{T}}_2^{n_2}\tilde{\mathsf{T}}_3^{n_3}]}}\nonumber\\
&&=\frac{(3+3^{n_3})(13+2\sqrt{2}\,3^{n_1}+3^{n_2}(3-\sqrt{2}+2\cdot3^{n_1}))}{4\sqrt{2}\sqrt{(5+3^{n_2}(1+2\cdot3^{n_1}))(3+3^{n_3})(3+3^{n_0})}}
\end{eqnarray}
using the configuration from Fig.~\ref{fig:mpssuppmat}a), where $n_0=n_1+n_2+n_3$,
\begin{eqnarray}
\mathsf{T}_1&=&\sum_{\sigma=0,\pm1}\mathsf{M}^*(\sigma)\otimes\mathsf{M}(\sigma)\otimes\mathbf{1}\otimes\mathbf{1}\\
\mathsf{T}_2&=&\sum_{\sigma=0,\pm1}
\mathsf{M}^*(\sigma)\otimes\mathbf{1}\otimes\mathsf{M}(\sigma)\otimes\mathbf{1}\\
\mathsf{T}_3&=&\sum_{\sigma=0,\pm1}
\mathsf{M}^*(\sigma)\otimes\mathbf{1}\otimes\mathbf{1}\otimes\mathsf{M}(\sigma)
\end{eqnarray}
and
\begin{eqnarray}
  \label{eq:38}
  \tilde{\mathsf{T}}_1&=&\sum_{\sigma=0,\pm1}\mathsf{M}^*(\sigma) \otimes\mathbf{1}\otimes\mathbf{1}\otimes\mathsf{M}(\sigma)\otimes\mathbf{1}\otimes\mathbf{1}\\
\tilde{\mathsf{T}}_2&=&\sum_{\sigma=0,\pm1}\mathbf{1}\otimes
\mathsf{M}^*(\sigma)\otimes\mathbf{1}\otimes\mathbf{1}\otimes\mathsf{M}(\sigma)\otimes\mathbf{1}\\
\tilde{\mathsf{T}}_3&=&\sum_{\sigma=0,\pm1}\mathbf{1}\otimes\mathbf{1}\otimes
\mathsf{M}^*(\sigma)\otimes\mathbf{1}\otimes\mathbf{1}\otimes\mathsf{M}(\sigma)
\end{eqnarray}
\item two loops connected to another two loops:
\begin{eqnarray}
    \label{eq:32}
   && \langle{\rm AKLT}_1,{\rm AKLT}_2|{\rm AKLT}_3,{\rm
      AKLT}_4\rangle\\
&&\qquad\qquad\qquad\qquad\qquad=\frac{{\rm
        Tr}[{\mathsf{T}_1'}^{n_3'}{\mathsf{T}_2'}^{n_1'-n_3'}{\mathsf{T}_3'}^{n_2'}]}{\sqrt{{\rm
        Tr}[{\tilde{\mathsf{T}}_1'}{}^{n_1'}{\tilde{\mathsf{T}}_2'}{}^{n_2'}]{\rm
        Tr}[{\tilde{\mathsf{T}}_1'}{}^{n_3'}{\tilde{\mathsf{T}}_2'}{}^{n_4'}]}}\nonumber\\
&&=\frac{(3+3^{n_2'})(13+2\sqrt{2}\,3^{n_3'}+3^{n_1'-n_3'}(3-\sqrt{2}+2\cdot3^{n_3'}))}{8\sqrt{(3+3^{n_3'})(3+3^{n_2'}) (3+3^{n_1'}) (3+3^{n_1'+n_2'-n_3'})}}\nonumber
\end{eqnarray}
using the configuration from Fig.~\ref{fig:mpssuppmat}b), with $n_1'+n_2'=n_3'+n_4'$, and where
\begin{eqnarray}
  \label{eq:39}
  \mathsf{T}_1'&=&\sum_{\sigma=0,\pm1}\mathsf{M}^*(\sigma)\otimes\mathsf{M}(\sigma)\otimes\mathbf{1}\otimes\mathbf{1}\\
\mathsf{T}_2'&=&\sum_{\sigma=0,\pm1}
\mathsf{M}^*(\sigma)\otimes\mathbf{1}\otimes\mathsf{M}(\sigma)\otimes\mathbf{1}\\
\mathsf{T}_3'&=&\sum_{\sigma=0,\pm1}
\mathbf{1}\otimes\mathbf{1}\otimes\mathsf{M}(\sigma)\otimes\mathsf{M}^*(\sigma)
\end{eqnarray}
and
\begin{eqnarray}
  \label{eq:40}
    \tilde{\mathsf{T}}'_1&=&\sum_{\sigma=0,\pm1}\mathsf{M}^*(\sigma) \otimes\mathbf{1}\otimes\mathsf{M}(\sigma)\otimes\mathbf{1}\\
\tilde{\mathsf{T}}'_2&=&\sum_{\sigma=0,\pm1}\mathbf{1}\otimes
\mathsf{M}^*(\sigma)\otimes\mathbf{1}\otimes\mathsf{M}(\sigma)
\end{eqnarray}
\end{itemize}
The overlaps Eqs.~\eqref{eq:31},\eqref{eq:32} take the form of a dominant $1/4$ contribution and
exponentially decaying terms. This (dominating) length-independent contribution is simply equal to
the overlap of ``neighboring'' $S=1/2$ spins from different chains
(see Fig.~\ref{fig:mpssuppmat}c)),
without the $S=1$ on-site projections, and may be empirically
understood from the very short range entanglement in the AKLT
wavefunction. In fact, $1/4$ is a special case of a more general
formula according to which the overlap of two coverings where $n_{\rm cuts}$ loop
cuts are needed to reconnect them is equal to $1/2^{n_{\rm
    cuts}-1}$. For example, if $|\psi_1\rangle$ and $|\psi_2\rangle$
are as depicted on Fig.~\ref{fig:mpssuppmat}c), then:
\begin{eqnarray}
  \label{eq:44}
&&|\psi_1\rangle=\frac{1}{2}\left(|\!\!\uparrow_1\downarrow_2\rangle-|\!\!\downarrow_1\uparrow_2\rangle\right)
   \left(|\!\!\uparrow_3\downarrow_4\rangle-|\!\!\downarrow_3\uparrow_4\rangle\right)\\
&&=\frac{1}{2}\left[|\!\!\uparrow_1 \downarrow_2\uparrow_3\downarrow_4\rangle+|\!\!\downarrow_1 \uparrow_2\downarrow_3\uparrow_4\rangle-|\!\!\uparrow_1 \downarrow_2\downarrow_3\uparrow_4\rangle-|\!\! \downarrow_1 \uparrow_2\uparrow_3\downarrow_4\rangle\right]\nonumber
\end{eqnarray}
and
\begin{eqnarray}
  \label{eq:43}
 &&\langle\psi_2|= \frac{1}{2}\left(\langle \uparrow_1\downarrow_4\!\!|-\langle
  \downarrow_1\uparrow_4\!\!|\right) \left(\langle
  \uparrow_3\downarrow_2\!\!|-\langle
  \downarrow_3\uparrow_2\!\!|\right)\\
&&=\frac{1}{2}\left[\langle \uparrow_1 \downarrow_2\uparrow_3\downarrow_4\!\!|+\langle \downarrow_1 \uparrow_2\downarrow_3\uparrow_4\!\!|-\langle \uparrow_1 \uparrow_2\downarrow_3\downarrow_4\!\!|-\langle \downarrow_1 \downarrow_2\uparrow_3\uparrow_4\!\!|\right]\nonumber
\end{eqnarray}
and so:
\begin{eqnarray}
  \label{eq:45}
\langle\psi_2|\psi_1\rangle=\frac{1}{2},
\end{eqnarray}
consistent with the fact that two loop cuts are needed to go from the
upper to the lower MPS, and vice versa.





\section{Details of the spin model phase diagram on the non-uniform honeycomb
  lattice}
\label{sec:details-spin-model}

Here we take a simple variational approach and calculate
the energy of three different states, for varying $\beta$ and
$\lambda_p$ for the pattern depicted in Fig.~\ref{fig:blbq}a). The
three states are the simple antiferromagnet, a valence bond solid
where the singlets lie on $\lambda_d$ bonds, and a Haldane hexagon
crystal where length-six Haldane chains form along $\lambda_p$ bonds. We consider the
Hamiltonian
\begin{eqnarray}
  \label{eq:33}
  H&=&\lambda_pJ\sum_{\langle ij\rangle\in
       p}\left(\mathbf{S}_i\cdot\mathbf{S}_j+\beta(\mathbf{S}_i\cdot\mathbf{S}_j)^2\right)\\
&&+\lambda_dJ\sum_{\langle
    ij\rangle\in d}\left(\mathbf{S}_i\cdot\mathbf{S}_j+\beta(\mathbf{S}_i\cdot\mathbf{S}_j)^2\right).
\end{eqnarray}
The approach could be
readily refined to the next simplest level of variational approach by
considering more general MPS network states (e.g.\ non-uniform along a
plaquette, still with bond dimension 2 to keep it simple),
but we deem it unnecessary for our purpose, as we see below. 

The energy per bond in the simple antiferromagnet, where the state of a bond is
of the form $|{\rm AFM}\rangle=|1-1\rangle$, is given by:
\begin{equation}
  \label{eq:12}
  \epsilon^{\rm AFM}(\lambda_p,\beta)=\frac{J}{3}(2\lambda_p+\lambda_d)(-1+2\beta)=\frac{2J}{3}(2\beta-1).
\end{equation}

The energy per bond in the valence bond solid with the valence bonds across
$\lambda_d$ bonds, the state of a $\lambda_d$ bond is $|{\rm VBS}\rangle=\frac{1}{\sqrt{2}}(|1-1\rangle-|-11\rangle)$ is
\begin{eqnarray}
  \label{eq:13}
  \epsilon^{\rm
  VBS}(\lambda_p,\beta)&=&\frac{J}{3}\left(-\lambda_d+\beta(\lambda_d+3\lambda_p)\right)\\
&=&\frac{2J}{3}\left(-1+\lambda_p+\beta(1+\frac{\lambda_p}{2})\right).
\end{eqnarray}

The energies of length-six Haldane chains at various $\beta$'s obtained
with exact diagonalization are given
Table~\ref{tab:length6beta}. To obtain the energy of the $\lambda_d$ bonds, we use the
eigenvectors, and find that it is simply proportional to $\beta$,
\begin{eqnarray}
  \label{eq:34}
  \epsilon^{\rm
  Haldane}(\lambda_p,\beta)&=&\frac{J}{3}\left(2\lambda_p\epsilon_6(\beta)+\beta\lambda_d\frac{4}{3}\right)\\
&=&\frac{2J}{3}\left(\frac{4\beta}{3}+\lambda_p(\epsilon_6(\beta)-\frac{4\beta}{3})\right).
\end{eqnarray}

Hence, the system is in the AFM phase if:
  \begin{equation}
    \label{eq:36}
  \begin{cases}
\lambda\geq\frac{2\beta}{2+\beta}\\
\lambda\leq\frac{3-2\beta}{4\beta-3\epsilon_6(\beta)}&\mbox{if }4\beta-3\epsilon_6(\beta)\geq0
\end{cases},
\end{equation}
the system is in the VBS phase if:
\begin{equation}
  \label{eq:35}
  \begin{cases}
\lambda\leq\frac{2\beta}{2+\beta}\\
\lambda\leq\frac{2(3+\beta)}{6+11\beta-6\epsilon_6(\beta)}&\mbox{if }6+11\beta-6\epsilon_6(\beta)\geq0
\end{cases},
\end{equation}
and the system is in the Haldane crystal phase if:
\begin{equation}
  \label{eq:37}
\begin{cases}
  \lambda\geq\frac{3-2\beta}{4\beta-3\epsilon_6(\beta)}&\mbox{if
  }4\beta-3\epsilon_6(\beta)\geq0\\
\lambda\geq\frac{2(3+\beta)}{6+11\beta-6\epsilon_6(\beta)}&\mbox{if
}6+11\beta-6\epsilon_6(\beta)\geq0
\end{cases}.
\end{equation}
Using these relations, we draw the approximate phase diagram shown in
Fig.~\ref{fig:blbq}b) in the main text. The results are consistent
with those obtained for the uniform exchange bilinear-biquadratic
model in Refs.~\onlinecite{zhao2012} and \onlinecite{corboz2013}.

\begin{figure}[htb]
  \centering
  \includegraphics[width=3in]{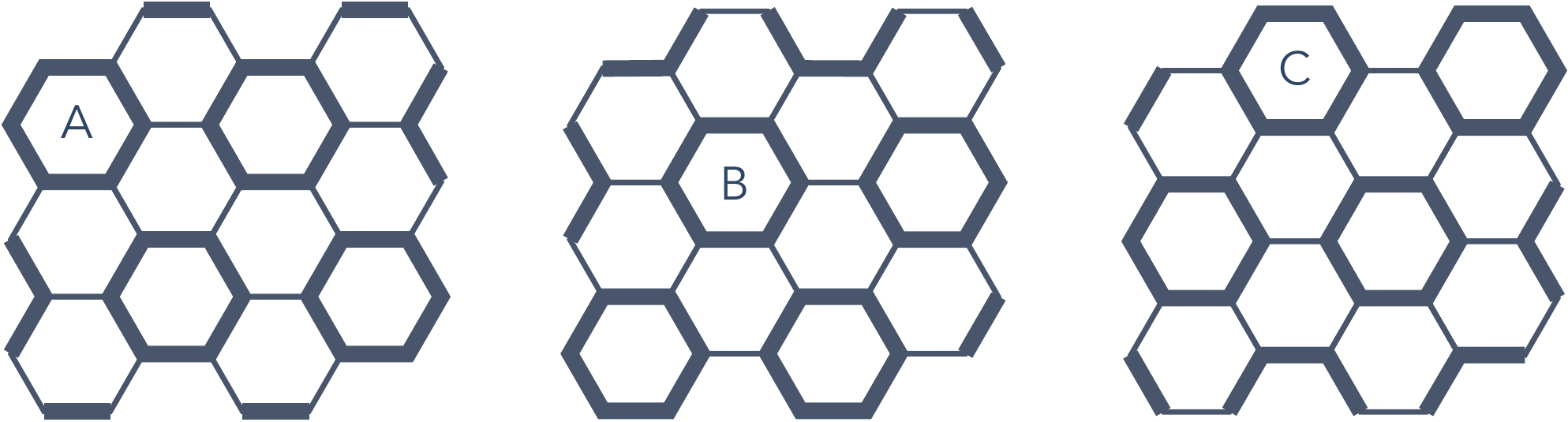}
  \caption{The three different hexagon loop crystals.}
  \label{fig:hextals}
\end{figure}

\section{Ground and excited state energies for chains of length six}
\label{sec:excitation-energies}

\begin{table}[h]
  \centering
  \begin{tabular}{c|ccccc}
\hline\hline
$\beta$ & $-1$ & $-2/3$ & $-1/3$ & $0$ & $1/12$ \\
$E_6^0$ & $-24.8774$ & $-19.2953$ & $-13.8305$ & $-8.61742$ & $-7.38496$\\
$E_6^1$ & $-22.6517$ & $-17.6728$ & $-12.7378$ & $-7.8968$ & $-6.71459$ \\
$\breve{E}_6^0$ & $-21.8484$ & $-16.901$ & $-12.0325$ & $-7.37027$ & $-6.2737$ \\
\hline\hline
$\beta$ & $1/6$ & $1/4$ & $1/3$ & $5/12$ & $1/2$ \\
$E_6^0$ & $-6.19548$ & $-5.06111$ & $-4$ & $-3.03786$ & $-2.1977$ \\
$E_6^1$ & $-5.55072$ & $-4.41121$ & $-3.30427$ & $-2.24068$ & $-1.27149$ \\
$\breve{E}_6^0$ & $-5.22356$ & $-4.2353$ & $-10/3$ & $-2.55428$ & $-1.91296$ \\
\hline\hline
$\beta$ & $7/12$ & $2/3$ & $3/4$ & $5/6$ & $11/12$ \\
$E_6^0$ & $-1.47434$ & $-0.834305$ & $-0.244138$ & $0.317211$ & $0.861294$ \\
$E_6^1$ & $-0.543207$ & $0.163247$ & $0.848938$ & $1.51519$ & $2.16354$ \\
$\breve{E}_6^0$ & $-1.36832$ & $-0.874155$ & $-0.406012$ & $0.0474453$
    & $0.491835$\\
\hline\hline
$\beta$ & $1$ \\
$E_6^0$ & $1.39445$ \\
$E_6^1$ & $2.7956$ \\
$\breve{E}_6^0$ & $0.930216$ \\
\hline\hline
  \end{tabular}
  \caption{Ground state energy, first excited state
    energy of periodic Haldane chains ($E_6^0$ and $E_6^1$, respectively) and ground state
    energy of open Haldane chains ($\breve{E}_6^0$) of length six as a function of
    $\beta$ (obtained in exact diagonalization), in units of $J$.}
  \label{tab:length6beta}
\end{table}

\end{document}